\documentclass{PoS}

\newcommand{\scaleFactor}{0.25}

\title{Search for the $^4\mbox{He}-\eta$ bound state with WASA-at-COSY}

\ShortTitle{Search for the $^4\mbox{He}-\eta$ bound state with WASA-at-COSY}

\author{\speaker{Wojciech KRZEMIEN}
\thanks{For the WASA-AT-COSY collaboration.}\\
        Jagiellonian University\\
        E-mail: \email{wojciech.krzemien@if.uj.edu.pl}
        }
        
\abstract{
We conduct a search for the ${^4\mbox{He}}-\eta$ bound state with WASA-at-COSY facility, via a measurement
of the excitation functions for the $dd \rightarrow {^3\mbox{He}} p \pi^{-}$ reaction, 
where the outgoing $p-\pi^{-}$ pairs originate from the conversion of the $\eta$ meson
on a nucleon inside the ${\mbox{He}}$ nucleus.
In June, 2008 first measurements of the excitation functions 
for the $dd \rightarrow {^3\mbox{He}}p\pi^{-}$ reaction were performed. 
In the experiment we used a slowly ramped COSY deuteron beam, scanning 
the range of momenta corresponding to the variation of the excess
energy for the $^4He-\eta$ system from  -51.4~MeV to 22~MeV.
}

\FullConference{8th International Conference on Nuclear Physics at Storage Rings-Stori11,\\
		October 9-14, 2011\\
		Laboratori Nazionali di Frascati dell'INFN, Italy}

\begin{document}

\section{Introduction}

The investigation of the exotic objects in the nuclear physics is a proven method for revealing many interesting properties of nuclear systems and for accessing to  an unexplored areas of physics. The recent progress in the spectroscopy of deeply bound pionic atoms  has permitted to obtain deeper insights into the meson-nucleus interaction and the in-medium behaviour of spontaneous chiral symmetry breaking~\cite{hiren}.

It is also conceivable that neutral mesons such as $\eta, \bar{K},\omega,\eta'$ can form bound states
with atomic nuclei.
In this case the binding is exclusively due to the strong interaction and the bound state - {\em mesic nucleus}
- can be considered as a meson moving in the mean field of the nucleons in the nucleus.
Due to the strong attractive $\eta$-nucleon interaction~\cite{wycech}, the $\eta$-mesic nuclei are ones of the most promising candidates for such states.

The discovery of the $\eta$-mesic nuclei would be interesting on its own but it would be also
valuable for investigations of the $\eta-N$ interaction and for the study of the in-medium properties
of the $N^*$ resonance~\cite{jido} and of the $\eta$ meson~\cite{osetNP710}.
It could also help to determine the flavor singlet
component of the $\eta$ wave function~\cite{basssymposium}.

The existence of $\eta$-mesic nuclei was postulated in 1986 by Haider and Liu \cite{liu2},
and since then a search for such states  was conducted in many experiments in the past~\cite{lampf,lpi,jinr,gsi,gem,mami} and is being continued at COSY~\cite{moskalsymposium,jurek-he3,timo,meson08}, JINR~\cite{jinr}, J-PARC~\cite{fujiokasymposium} and MAMI~\cite{mami}.
However, up to now no firm experimental evidence for $\eta$-mesic nuclei was found.

A very strong final state interaction (FSI) observed
in the $dd \rightarrow {^4\mbox{He}} \eta$ reaction close to kinematical threshold
and interpreted as possible indication of ${^4\mbox{He}}-\eta$ bound state~\cite{Willis97}
suggests, that ${^4\mbox{He}}-\eta$ system is a good candidate for experimental study of possible binding.

Taking into account the above arguments, we performed a search for $\eta$-mesic $^{4}\mbox{He}$
by measuring the excitation function for the $dd \rightarrow {^3\mbox{He}} p\pi^-$  reaction
in the vicinity of the $\eta$ production threshold.

%Many promising indications where reported, however, so far there is no direct experimental confirmation of the existence of mesic nucleus. In the region of the light nuclei systems such as $\eta$-$\mbox{He}$ or $\eta$-$\mbox{T}$, the observation of the strong enhancement in the total cross-section and the phase variation in the close-to-threshold region provided strong evidence to the hypothesis of the existence of a pole in the scattering matrix which can correspond to the bound state~\cite{wilkin}.

%However, as it was stated in ~\cite{liu3,haider2}, the theoretical predictions of width and binding energy of the $\eta$-mesic nuclei is strongly dependent on the not well known  subtreshold $\eta$-nucleon interaction. Therefore, the direct measurements which could confirm the existence of the bound state, are mandatory.
\section{Method}
\begin{figure}[!ht]
\begin{center}
      \scalebox{0.5}
         {
              \includegraphics{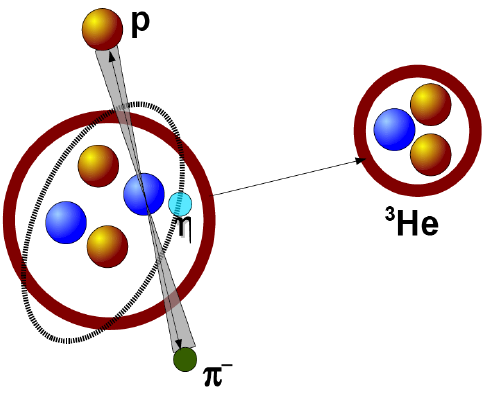}
         }
\caption[Bound state decay scheme]{\label{decay_scheme_p}Schematic picture of the  $({^4\mbox{He}}-\eta)_{bound} \rightarrow {^3\mbox{He}} p \pi^{-}$
decay. In the first step the $\eta $ meson is absorbed on one of the neutrons and the $N^{\star} $ resonance is formed. Next, the N$^{\star}$ decays into a $p-\pi^{-}$ pair. The $^3\mbox{He}$ plays the role of a spectator. }
\end{center}
\end{figure}

In our experimental studies, we used the deuteron-deuteron collisions at energies around the $\eta$ production threshold for production of the $\eta-{^4\mbox{He}}$ bound state.
We expect, that the decay of such state proceeds via absorption of the $\eta$ meson on one of the nucleons
in the ${^4\mbox{He}}$ nucleus leading to excitation of the  $N^{\star}$(1535)  resonance which subsequently
decays in pion-nucleon pair.
The remaining three nucleons play a role of spectators and they are likely to bind forming ${^3\mbox{He}}$
or ${^3\mbox{H}}$ nucleus.
This scenario is schematically presented in Fig.~\ref{decay_scheme_p}.

According to the discussed scheme, there exist four equivalent decay channels
of the $({^4\mbox{He}}-\eta)_{bound}$ state:
\begin{itemize}
\item   $({^4\mbox{He}}-\eta)_{bound} \rightarrow {^3\mbox{He}} p \pi^{-}$
\item   $({^4\mbox{He}}-\eta)_{bound} \rightarrow {^3\mbox{He}} n \pi^{0}$
\item   $({^4\mbox{He}}-\eta)_{bound} \rightarrow {^3\mbox{H}} p \pi^{0}$
\item   $({^4\mbox{He}}-\eta)_{bound} \rightarrow {^3\mbox{H}} n \pi^{+}$
\end{itemize}

\begin{figure}[!ht]
\begin{center}
      \scalebox{\scaleFactor}
         {
              \includegraphics{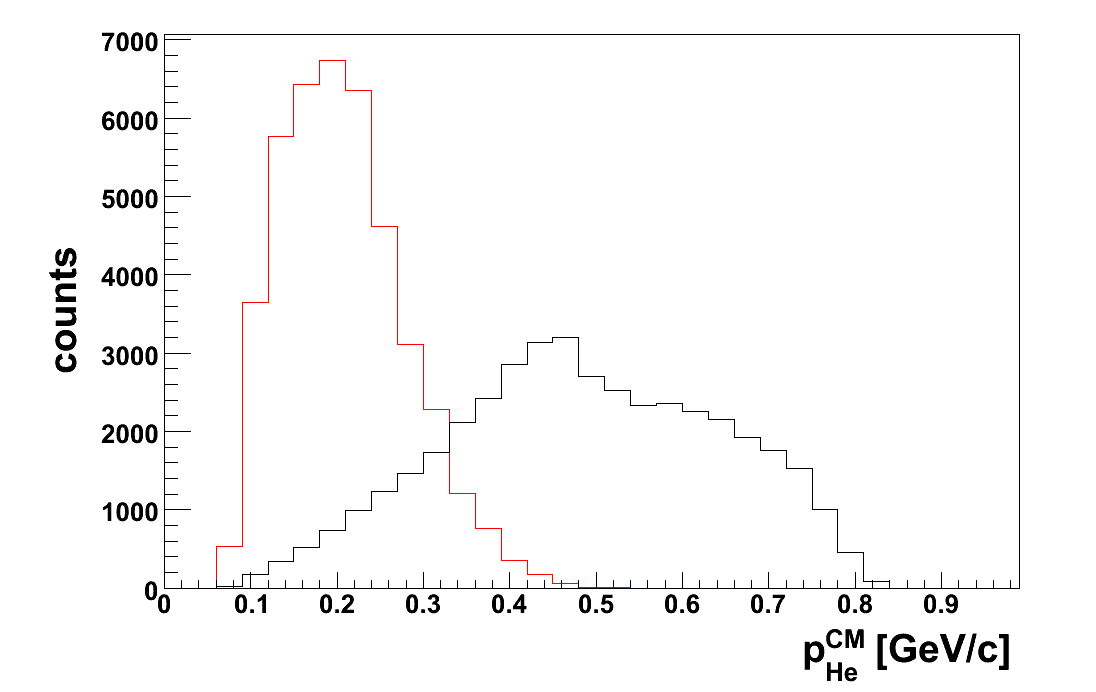}
         }

\caption[Distribution of the ${^3\mbox{He}}$ momentum in CM]{\label{pmomCM_p} Distribution of the ${^3\mbox{He}}$ momentum in the CM system obtained in simulation of the processes leading to the creation
of the  ${^4\mbox{He}}\eta$ bound state:
$dd \rightarrow (^4{\mbox{He}} \eta)_{bound} \rightarrow {^3{\mbox{He}}}p\pi^{-}$ (red line)
and of the direct $dd \rightarrow {^3{\mbox{He}}} p\pi^{-}$ decay (black line).
%The simulation was done for momentum of the deuteron beam of 2.307~GeV/c.
}
\end{center}
\end{figure}

In our experiment we concentrated on the ${^3\mbox{He}} p \pi^{-}$ decay mode
due to the highest acceptance of the WASA-at-COSY detector in this case.
The outgoing $^3\mbox{He}$ nucleus plays the role of a spectator and, therefore,
we expect that its momentum in the CM frame is relatively low and can be described
by the Fermi momentum distribution of nucleons in the $^4\mbox{He}$ nucleus.
This signature  allows to suppress background from reactions leading to the
${^3\mbox{He}} p \pi^{-}$ final state but proceeding without formation of the intermediate
$({^4\mbox{He}}-\eta)_{bound}$ state and, therefore, resulting on the average in much higher
CM momenta of $^3\mbox{He}$ (see Fig.~\ref{pmomCM_p}).

\begin{figure}[!ht]
\begin{center}
      \scalebox{\scaleFactor}
         {
              \includegraphics{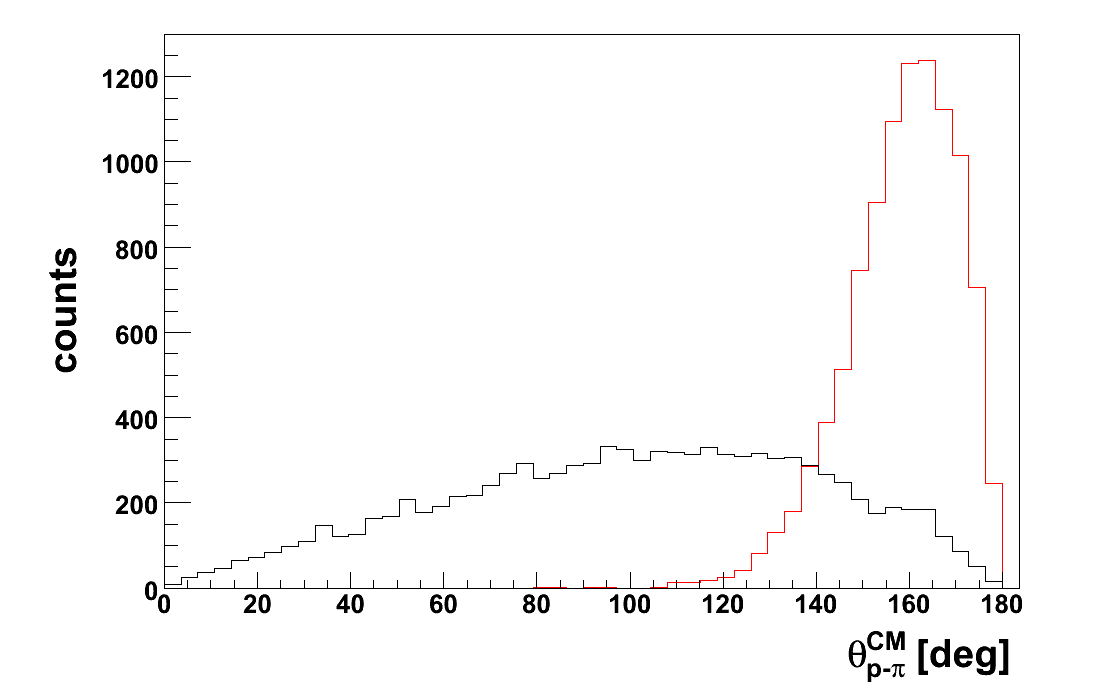}
         }

\caption[Distribution of the relative $p-\pi^{-}$ angle seen in CM]{\label{proton_pion_openAngleCM_p}
Distribution of the  $p-\pi^{-}$ opening angle in the CM system obtained in simulation of the processes
leading to the creation of the ${^4\mbox{He}}\eta$ bound state:
$dd\to({^4\mbox{He}}\eta)_{bound}\to{^3\mbox{He}}p\pi^{-}$ (red line) and
of the direct $dd\to{^3{\mbox{He}}}p\pi^{-}$ reaction (black line).
%The simulation was done for momentum of the deuteron beam of 2.307~GeV/c.
}
\end{center}
\end{figure}

A kinematic variable correlated with the $^3\mbox{He}$ CM momentum is the relative angle  of the outgoing
$p-\pi^{-}$ pair.
In the limit of $^3\mbox{He}$ produced at rest in the CM frame this angle is exactly equal to ${180^{\circ}}$
but due to the presence of the Fermi motion it is smeared by about ${30^{\circ}}$(see Fig.~\ref{proton_pion_openAngleCM_p}).

The principle of the present experiment was based on the measurement of the excitation function
of the $dd \rightarrow {^3\mbox{He}} p \pi^{-}$ reaction for energies in the vicinity of the $\eta$ production
threshold and on the selection of events with low ${^3\mbox{He}}$ CM momenta.
In the case of existence of the ${^4\mbox{He}}-\eta$ bound state we expected to observe
a resonance-like structure in the excitation function at the reaction CM energies below the $\eta$ threshold.
From the central energy of the observed structure $E^{CM}$ one can determine the binding energy
of the $({^4\mbox{He}}-\eta)$ system:
\begin{equation}
E_{BE}=m_{He}+m_{\eta} - E^{CM}.
\end{equation}
The width of the structure is equal to the width of the bound state.

\section{Experiment}

In June 2008 we performed a search for the $\eta$-mesic ${^4\mbox{He}}$ by measuring the excitation fun\-ction of the $dd \rightarrow ^3\mbox{He} p\pi^-$  reaction near the $\eta$ meson production threshold using the WASA-at-COSY detector~\cite{jurekmeson08}. During the experimental run the momentum of the deuteron beam was varied continuously within each acceleration cycle
from  2.185~GeV/c to 2.400~GeV/c, crossing the kinematic threshold for the $\eta$ production in the $dd \rightarrow {^4\mbox{He}}\,\eta$ reaction at 2.336~GeV/c.
This range of beam momenta corresponds to the variation of $^4\mbox{He}-\eta$  excess energy  from -51.4~MeV to 22~MeV.

The identification of the $^{3}{\mbox{He}}$ was conducted using the  $\Delta E- \Delta E$ techniques comparing the energy losses in two layers of the Forward Range Hodoscope. The energy loss in the Plastic Scintillator Barrel was combined with the energy deposited in the Electromagnetic Calorimeter to identify protons and pions.

\begin{figure}[!hp]
\begin{center}
  \scalebox{\scaleFactor}
  {
    \includegraphics{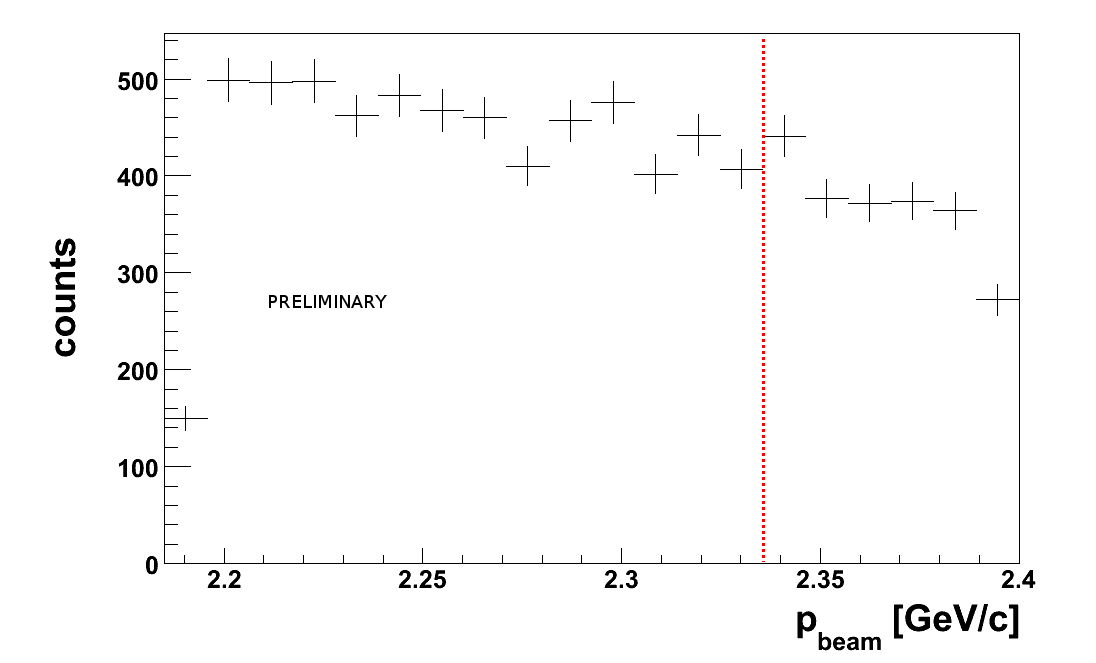}
  }
  \scalebox{\scaleFactor}
  {
    \includegraphics{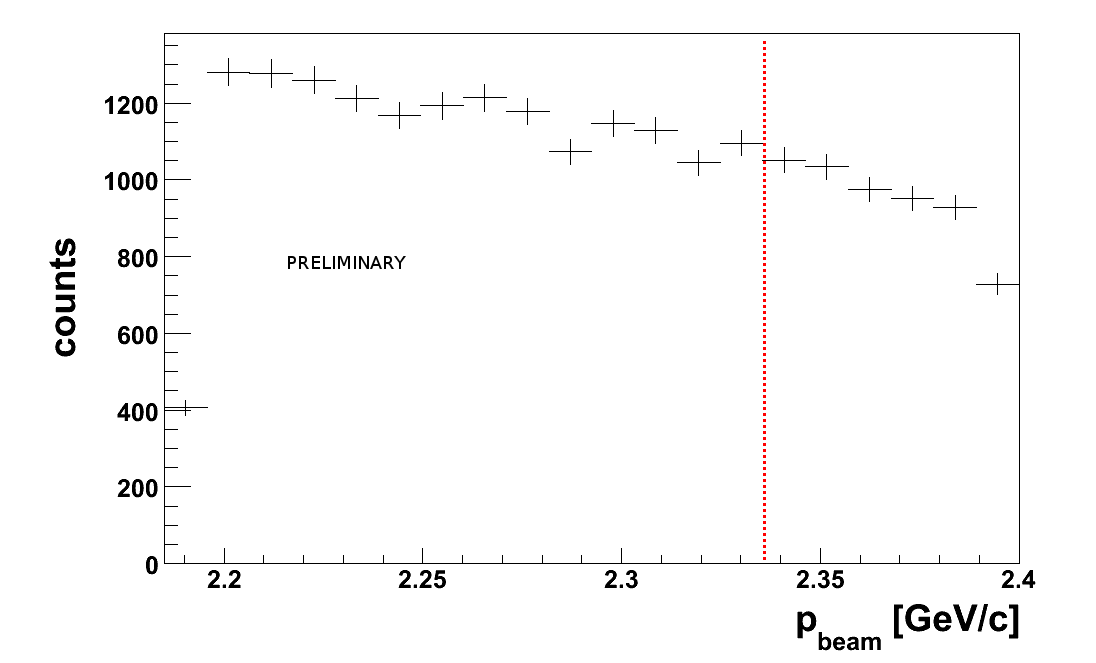}
  }
  \scalebox{\scaleFactor}
  {
    \includegraphics{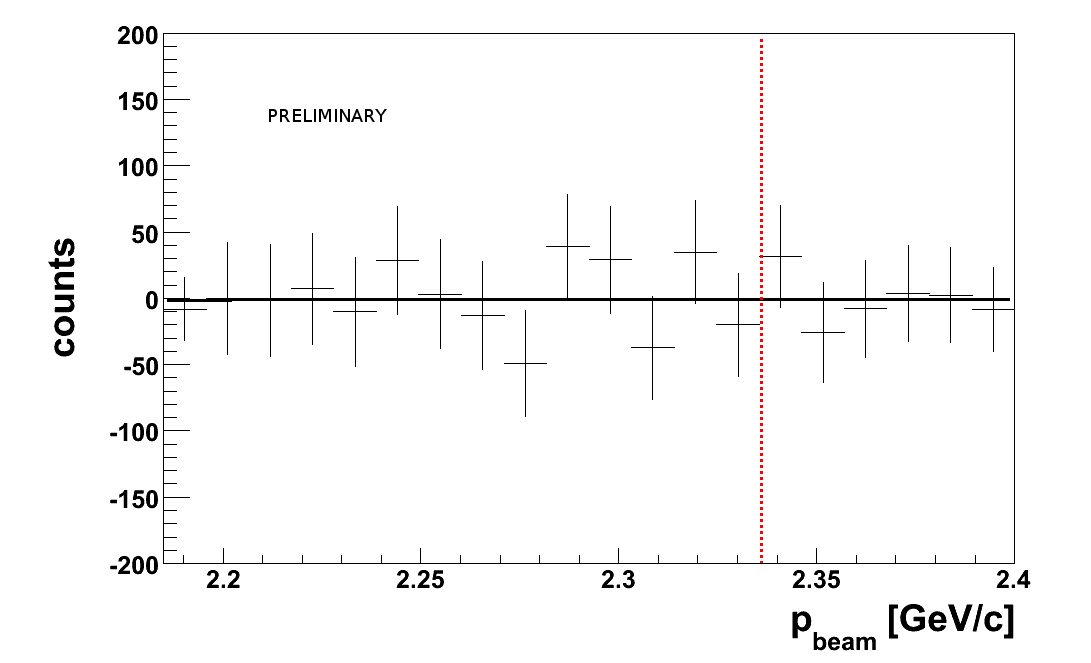}
  }
\caption[Not normalized excitation functions]{\label{hExcitFuncCM_mom_exp_bad_p}Excitation function for the $dd \rightarrow {^3\mbox{He}} p \pi^{-}$ reaction for the "signal-rich" region corresponding to ${^3\mbox{He}}$ momentum below 0.3\,GeV/c (upper panel) and the "signal-poor" region with  ${^3\mbox{He}}$ momentum above 0.3\,GeV/c (middle panel).
Difference of the excitation functions for the "signal-rich" and "signal-poor" regions after the normalization
to the lowest beam momentum bin is shown in the lower panel. The black solid line represents a straight line fit.
The beam momentum corresponding to the ${^4\mbox{He}}-\eta$ threshold is marked by the vertical red line.}
\end{center}
\end{figure}

We constructed two types of excitation function for the $dd \rightarrow {^3\mbox{He}} p \pi^{-}$ reaction.
They differ in the selection of the events and in the way of normalizing the data points.
The first excitation function uses events from the  "signal-rich" region
corresponding to the ${^3\mbox{He}}$ CM momenta below 0.3\,GeV/c. 
The counts are plotted as a function of the beam momentum as it is shown Fig.~\ref{hExcitFuncCM_mom_exp_bad_p}(top).
The obtained function is smooth an no clear signal, which could be interpreted as a resonance-like
structure, is visible.
A similar dependence was obtained for events originating from the "signal-poor" region
corresponding to ${^3\mbox{He}}$ CM momenta above 0.3\,GeV/c (see Fig.~\ref{hExcitFuncCM_mom_exp_bad_p}(middle)).
We checked also for possible structures in the difference between the discussed functions
for the "signal-rich" and "signal-poor" region.
We multiplied the function for the "signal-poor" region by a factor chosen in such a way,
that the difference of the two functions for the lowest beam momentum bin is equal to zero.
This difference is presented in Fig.~\ref{hExcitFuncCM_mom_exp_bad_p}(bottom).
The obtained dependence is flat and is consistent with zero. No resonance structure is visible.

\begin{figure}[!ht]
\begin{center}
      \scalebox{\scaleFactor}
      {
         \includegraphics{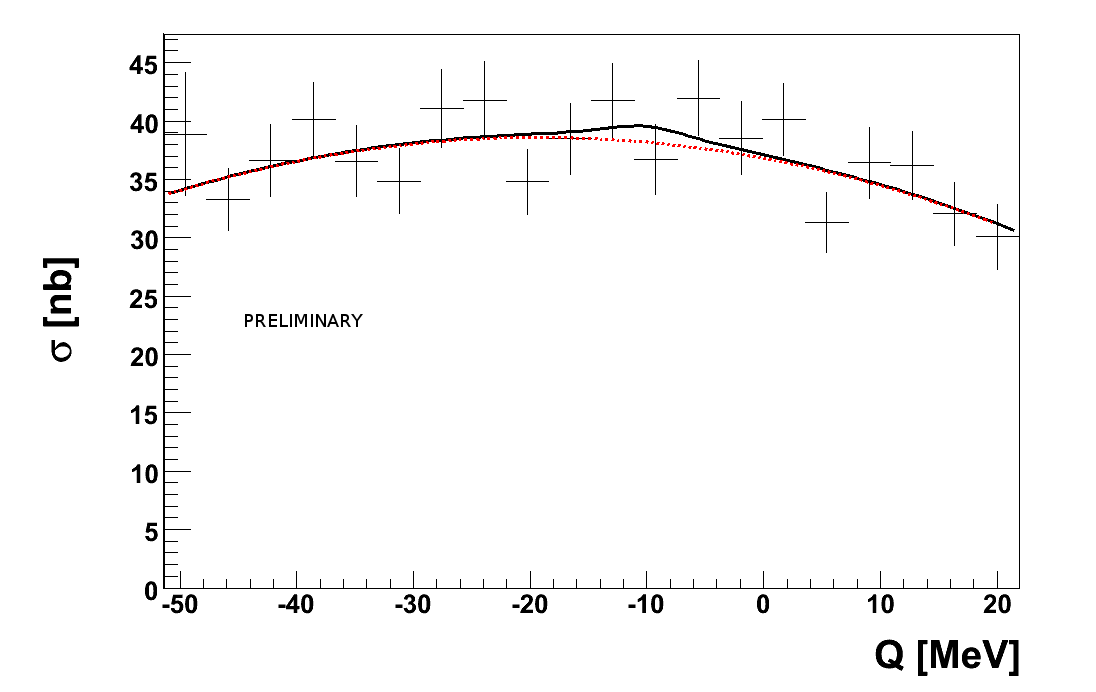}
      }
\caption[Breit-Wigner fit to the normalized excitation function]{\label{fit_10_10_allcuts_p} 
Excitation function for the $dd \rightarrow {^3\mbox{He}} p \pi^{-}$ reaction obtained by normalizing the events
selected in individual excess energy intervals by the corresponding integrated luminosities.
The solid line represents a fit with second order polynomial combined with a Breit-Wigner function 
with  fixed binding energy and width equal to -10 and 10 MeV, respectively. 
The dotted line corresponds to the contribution from the second order polynomial in the performed fit. The $\sigma$ values are not corrected for acceptance and efficiency cuts. }
\end{center}
\end{figure}

However, we do not treat this result as a final conclusion of non observation
of the ${^3\mbox{He}}-\eta$ bound state,
since one can apply further cuts to reduce the background.
Additional cuts on the $p$ and $\pi^-$ kinetic energy distributions
and the $p-\pi^-$ opening angle in the CM system lead us to the construction of a second excitation curve.
The CM kinetic energies of protons and pions originate
from the mass deficit $m_{\eta}-m_{\pi}$ are around 50\,MeV and 350\,MeV, respectively.
We selected the kinetic energy of protons smaller than 200 MeV  and of pions from the interval (180, 400) MeV.
We applied also a cut on the relative $p-\pi^-$ angle in the CM system in the range of (140$^{\circ}$-180$^{\circ}$).
The number of selected events in each beam momentum interval was divided
by the corresponding integrated luminosity. This result is shown in Fig.~\ref{fit_10_10_allcuts_p}.

The integrated luminosity in the experiment was determined using the $dd \rightarrow {^3\mbox{He}} n$ reaction. 
%and it equals 117.9 $\pm$13.6\,$nb^{-1}$~\cite{wkPhD}.
The relative normalization of points of the $dd \rightarrow {^3\mbox{He}} p \pi^-$ excitation
function was based on the quasi-elastic proton-proton scattering.

In order to use the Breit-Wigner distribution for the description of a possible resonance structure
in the excitation function, we translated the beam momentum intervals into intervals of the excess energy with respect
to the ${^4\mbox{He}} \eta$ production threshold.
The excitation function presented in Fig.~\ref{fit_10_10_allcuts_p}
 can be  well described by a second order polynomial resulting in the chi-squared value per degree of freedom of 0.98.
%chi-square=24.05 n.d.f=liczba_obserwacji-liczba_fitowanych_parametrow-1=20-3-1=16
In the excitation function we observe no structure which could be interpreted
as a resonance originating from decay of the $\eta$-mesic ${^4\mbox{He}}$.

We assumed, that a signal from the bound state in the excitation curve  determined as a function of the excess energy $Q$ 
with respect to the ${^4\mbox{He}}-\eta$ threshold, can be described by the Breit-Wigner shape:
\begin{equation} \label{bw_eq}
\sigma(Q,E_{BE},\Gamma,A)=\frac{A \cdot (\frac{\Gamma}{2})^2}{(Q-E_{BE})^2 +(\frac{\Gamma}{2})^2},
\end{equation}
where $E_{BE}$ is the binding energy,  $\Gamma$ is the width and $A$ is the amplitude.
The value of the Breit-Wigner function for the central energy ($Q=E_{BE}$)  corresponds to the maximum cross-section
for the decay of the  $\eta$-mesic ${^4\mbox{He}}$ into the ${^3\mbox{He}} p \pi^{-}$ channel.
In order to determine an upper limit for the cross-section for formation of the ${^4\mbox{He}}-\eta$ bound state and its decay into the ${^3\mbox{He}} p \pi^{-}$ channel  we fitted the excitation function 
with quadratic function describing the background combined with the Breit-Wigner function.
In the fit we adjusted the quadratic background and the amplitude $A$ of the Breit-Wigner distribution.
The binding energy $E_{BE}$ and the width $\Gamma$ were fixed during the fit.
An example of the fit with $E_{BE}$=-10\,MeV and $\Gamma$=10\,MeV is shown in Fig.~\ref{fit_10_10_allcuts_p}.
The fit was performed for various values of the binding energy and the width representing different hypothesis of the bound state
properties. In each case, the value of the amplitude $A$ is consistent with zero within the uncertainty $\sigma_{A}$, 
which confirms the  hypothesis of non-observation of the signal.

%Obtained geometrical acceptance is about 60\% and the full efficiency including all cuts applied in the analysis is about 19\% and it varies by only about 1\% along the whole beam momentum range.

%In order to  calculate  an upper limit for the $dd \to (^4\mbox{He}\eta)_{bound} \rightarrow {^3\mbox{He}} p \pi^-$
%cross-section, the $\sigma_{A}$ values obtained in the above described fit
%had to be corrected for the efficiency $\varepsilon$ (equal to 19\%) and multiplied by the statistical factor $k$ equal to 1.64485 corresponding to the probability confidence level of 90\%:

%We obtained the preliminary upper limits for the cross-sections of 28, 32 and 41\,nb for production and decays of the bound state 
%with a width of 10, 20 and 30\,MeV, respectively.

\section{Outlook}

In November 2010 a new two-week measurement was performed with WASA-at-COSY. We collected  data with approximately 20 times higher statistics. In addition to the $dd \rightarrow {^3\mbox{He}} p \pi^-$ channel we registered also the $dd \rightarrow {^3\mbox{He}} n \pi^{0}$ reaction. The data analysis is undergoing. After two weeks of measurement with an estimated luminosity of $4 \cdot 10^{30}$ cm$^{-2}$ s$^{-1}$, we expect a statistical sensitivity of a few nb ($\sigma$). A non-observation of this signal will significantly lower the upper limit for the existence of the bound state. 

\section{Support}
This work has been supported by FFE funds of Forschungszetrum Juelich, grant No  41831803 (COSY-107), by the European Commission under the 7th Framework Programme through the 'Research Infrastructures' action of the 'Capacities' Programme. Call: FP7-INFRASTRUCTURES-2008-1, Grant Agreement N. 227431 and by the Polish Ministry of Science and Higher Education under grants No. 2367/B/H03/2009/37 and 0320/B/H03/2011/40.

\end{document}